\documentstyle[twocolumn,prb,aps]{revtex}
\newcount\equationno
\global\equationno=0
\def\EQNO{\global\advance\equationno by 1 \eqno(\the\equationno) }
\def\l{\hbox to \hsize}
\def\m#1{\hfill}
\def\undertext#1{$\underline{\smash{\hbox{#1}}}$}
\def\uncatcodespecials{\def\do##1{\catcode`##1=12 }\dospecials}
{\obeyspaces\global\let =\ } 
{\catcode`\`=\active \gdef`{\relax\lq}}
\def\v{\begingroup\hskip\parskip\tt\uncatcodespecials
  \obeyspaces\doverbatt}
\newcount\balance
{\catcode`<=1 \catcode`>=2 \catcode`\{=12 \catcode`\}=12
  \gdef\doverbatt`<\balance=1\verbatimloop>
  \gdef\verbatimloop#1<\def\next<#1\verbatimloop>%
    \if#1`\advance\balance by-1
     \ifnum\balance=0\let\next=\endgroup\fi\fi\next>>
\begin{document}
\title{CLUSTER MONTE CARLO ALGORITHMS AND THEIR APPLICATIONS$^*$}

\author{\it Jian-Sheng Wang}

\address{\
\\{\rm Computational Science Program,
\\National University of Singapore,
\\Lower Kent Ridge Road, Singapore 119260,
\\Republic of Singapore}
\smallskip
\\cscwjs@leonis.nus.sg
\\ \ }

\maketitle
\begin{abstract}
We review the background of the cluster algorithms in Monte Carlo
simulation of statistical physics problems.  One of the first such
successful algorithm was developed by Swendsen and Wang eight years
ago.  In contrast to the local algorithms, cluster algorithms update
dynamical variables in a global fashion.  Therefore, large changes are
made in a single step.  The method is very efficient, especially near
the critical point of a second-order phase transition.  Studies of
various statistical mechanics models and some generalizations of the
algorithm will be briefly reviewed.  We mention applications in other
fields, especially in imaging processing.
\end{abstract}


\section{Monte Carlo Method}
One of the important task in statistical mechanics and quantum field
theory is to compute the statistical average,
$$ \langle A \rangle = { \sum\limits_X A(X) \exp[-\beta H(X)] \over
\sum\limits_X \exp[-\beta H(X)] } =\!\! \sum_X \! A(X) P(X), \EQNO$$
where the summation is over a set of all states $X$, which usually is very
large.  A general method was proposed long time ago in the fifties by
Metropolis et al \cite{Metropolis,Kalos}, well-known as Metropolis
algorithm.  Instead of summing over all the states as required, one
generates a selection of the states according to the probability
$P(X)$, so that the answer is given approximately by an arithmetic
average of the quantity $A(X)$.  The generation of a sequence of the
states is done stochastically by the method of Markov chain.

\subsection{Markov chain}
A Markov chain \cite{Reichl} is a mathematical notion for a
random walk in the state
space $X$.  We start from some initial point $X_0$, the next point
$X_1$ is generated according to certain transition probability.  In
this way a sequence of random points $X_0,\, X_1,\, \cdots,\, X_N$ is
generated.  These points are required to appear with the probability
$P(X)$.  The rule to generate next point $X_{i+1}$ given the point
$X_i$ is specified by the transition probability
$$P(X_{i+1} | X_i) = W(X_i \to X_{i+1}). \EQNO $$
where $P(X_{i+1}| X_i)$ is the probability that the system is in state
$X_{i+1}$, given that the system was in state $X_i$, namely, it is the
conditional probability.  We have
$$ W(X_i \to X_{i+1}) \geq 0, \quad
\sum_{X_{i+1}} W(X_i \to X_{i+1}) = 1.\EQNO $$
Since $W$ is probability, it satisfies the usual constraint of
probability---probability must be positive and total probability adds
to one.

If we choose $W$ so that
$$ P(X) = \sum_{X'} P(X') W(X' \to X), \EQNO $$
we say that $P(X)$ is invariant under the transition of $W$, or $P(X)$
is the equilibrium distribution for the given transition probability.
Such a probability distribution will be unique if starting from any
state $X$, it is possible to get to any another state $X'$ after a
finite number of transitions.  We call such Markov chain ergodic
(regular, or irreducible).

\subsection{Detailed balance}
How to choose $W$ to get prescribed $P$?  It is sufficient that $W$
satisfies
$$ P(X') W(X'\to X) = P(X) W(X\to X'), \EQNO$$
where $P$ is the known distribution.  If $W$ satisfies this equation
and it is ergodic, then the Markov chain will generate $X$ according
to the probability distribution $P(X)$.

\subsection{Metropolis algorithm}
Metropolis method goes as follows:
starting from some initial state (configuration)
$X_0$, a sequence of states will be generated by a Markov chain
through the transition probability $W(X_i \to X_{i+1})$.  The
starting state may be a unique state or may be generated at random
with certain distribution $P_0$.  The new state $X_{i+1}$ is based on
the old state $X_i$.  The new state is generated in such a way so
that the probability that the system is in state $X_{i+1}$ is just
$W(X_i \to X_{i+1})$, knowing that the system was in state $X_i$.
Hopefully we can generate this probability distribution easily.
Otherwise we might have generated $P(X)$ directly in the first place.
It is sufficient that $W$ satisfies detailed balance condition
and the Markov chain is ergodic.
Metropolis algorithm refers to a specific choice of $W$:
$$ W(X \to X') = T(X\to X') \min\left( 1, {P(X')\over P(X)} \right), \EQNO$$
where $X \neq X'$ and $T(X\to X') = T(X'\to X)$.
Matrix $T$ can be any distribution but must be symmetric.
We almost always use
$$ T(X\to X') = \cases{ const, &
		for $X'$ in a region around $X$; \cr
                       0, & otherwise. \cr } \EQNO$$
The probability that the state does not change,
$ W(X\to X)$, is determined by the normalization condition
$$ \sum_{X'} W(X \to X') = 1.\EQNO$$

\section{Ising Model for Magnets}
To be able to understand Swendsen-Wang and other cluster algorithms,
we need to introduce the very popular model for ferromagnets in
condensed matter physics.  Ising model \cite{Ising,Onsager} is the
simplest model for ferromagnets.  One of the striking feature of a
magnet is that it has spontaneous magnetization at lower temperatures.
However, the magnetization disappears at some higher temperature
$T_c$.  The phase below $T_c$ is called ferromagnetic phase and that
above $T_c$ is called paramagnetic phase. $T_c$ is called critical
temperature and there is a phase transition at this temperature.

Let's take two-dimensional system as an example.  On an $L \times L$
square lattice, we put a spin $\sigma_i$ at each lattice site $i$.
Each spin takes only two possible values $+1$ and $-1$.  The set of
$L^2$ spins consists of the state space $X$.  The system has a total
energy as a function of the state $\{ \sigma \}$, defined by
$$ H\bigl(\{\sigma\}\bigr) =
- J \sum_{\langle i,\, j\rangle} \sigma_i \sigma_j, \EQNO $$
where $J$ is called coupling constant.
The summation is over the nearest neighbors only, i.e.,
each bond is summed once.  Such model is also referred as nearest
neighbor Ising model.
When $J$ is positive, the model describes a ferromagnet; while
$J<0$ describes an anti-ferromagnet.   Let's consider just one term in
the nearest neighbor interaction.    The interaction energy of neighbors
takes only two values, $+J$ or $-J$.  For ferromagnet, the energy
is lower if the two spins are parallel.  It costs energy
if the spins are pointing in different directions.

According to statistical-mechanical description, the system will not be
in a definite state.  Due to thermal fluctuation, the system will have
some probability distribution among the states.  If the temperature
$T$ is a fixed parameter, we have the famous Boltzmann distribution
$$ P\bigl(\{ \sigma\}\bigr) \propto
\exp\left(-{ H(\{ \sigma\}) \over k_B T } \right),\EQNO$$
where $k_B \approx 1.38 \times 10^{-23}\,$Joule$/$Kelvin is called the
Boltzmann constant.  Our goal is to calculate average quantities such
as energy, magnetization, correlation functions, etc. The internal
energy is defined by
$$ U =  \langle H \rangle =
\sum_{\{\sigma\}} H\bigl(\{\sigma\}\bigr)
P\bigl(\{\sigma\}\bigr); \EQNO$$
and the magnetization is defined as
$$  \langle M \rangle =
 \sum_{\{\sigma\}} \Bigl|\sum_i\sigma_i\Bigr|
 P\bigl(\{\sigma\}\bigr). \EQNO$$
The summation is over all possible states.  Since each site can have
two states (spin up and spin down), we have $2^{L^d}$ states for a
system of linear size $L$ in $d$ dimensions.

A Monte Carlo simulation of the Ising model by Metropolis algorithm
involves the following steps:
\begin{enumerate}
\item Initialize $\sigma_i$ at each site with arbitrary spin values,
e.g., all spin up, or up or down with equal probability.
The complete set of spins $\{\sigma\}$ is the abstract state $X$
discussed in the previous section.
\item Choose a site $k$ at random and propose to flip the spin at that
site, $\sigma'_k = - \sigma_k$.  The proposed state $\overline X$ is
the one with one spin at location $k$ flipped.  This amounts to take
$T( X\!\!\! \to\!\!\! \overline X)$ equal to $1/L^d$ if $X$ and
$\overline X$ differ by 1 spin, and 0 otherwise.
\item Calculate the energy increment
$$ \Delta E = H\bigl(\{\sigma'\}\bigr) - H\bigl(\{\sigma\}\bigr)
  = 2 J \sigma_k\!\!\!\! \sum_{ {\rm neighbors} \atop {\rm of\ } k } \!\!\!
\sigma_i. \EQNO$$
\item Accept the proposed state as
the new state if a uniformly distributed random number (between 0 and 1)
is less than $\exp(-\Delta E/k_B T)$; retain the old state as the new
state otherwise.
\item go to 2.
\end{enumerate}
\smallbreak

One Monte Carlo step will be defined as performing the above basic
single-spin flip $L^d$ number of times.  After one Monte Carlo step,
each site on average has tried once to flip.

\section{Critical Slowing Down}
In Ising model simulation, or more generally any simulation using the
Metropolis algorithm, the next configuration depends on the previous
one.  Due to this correlation between Monte Carlo steps, the formula
for the statistical error $\epsilon \approx \sigma / \sqrt{N}$
underestimates the true error, where $\sigma$ is the variance of the
quantity of interest, and $N$ is Monte Carlo steps.  The error formula
should be replaced by
$$ \epsilon \approx {\sigma \sqrt{ {1+2\tau \over N}} } .\EQNO $$
That is, the error is larger by a factor $\sqrt{1 + 2\tau}$ than the
uncorrelated data.  The quantity $\tau$ is called correlation
time, which is roughly the number of Monte Carlo steps needed to
generate independent configurations.  We can compute the correlation
time $\tau$ as follows: define the time-dependent correlation function,
$$ f(t) = { \langle A(t') A(t'+t) \rangle - \langle A\rangle^2
\over \langle A^2 \rangle - \langle A\rangle^2 }, \EQNO $$
where time $t$ is measured in terms of Monte Carlo steps---time is
passed by one unit when one updates the system by one Monte Carlo
step.  And $A(t')$ is the quantity at steps $t'$ whose error we want
to estimate.  The angular brackets denote average over Monte Carlo steps.
The correlation time is defined by
$$ \tau = \sum_{i=1}^\infty f(t). \EQNO $$

Now we are ready to explain the concept of critical slowing down.  Just
like many other quantities near the critical point, the correlation time
also becomes singular at the critical point.  On an infinite lattice,
it diverges with a power law
$$\tau \propto | T - T_c |^{-\nu z}. \EQNO$$
Here $\nu$ is the correlation length exponent and $z$ is called
dynamic critical exponent.  This phenomena that the correlation time
becomes very large is called critical slowing down.  It not only
happens in computer simulation but also appears in real systems.  On a
finite system of linear size $L$, the correlation time will, of
course, not go to infinity.  But it will grow with size as
$$ \tau \propto L^z,\quad T = T_c. \EQNO $$
Substituting this result into the error formula, we find that
$\epsilon \approx \sigma L^{z/2}/N^{1/2}$ for large system near $T_c$.
For the two-dimensional Ising model with Metropolis algorithm, $z
\approx 2.1$, we see typically that increasing the size leads to a
larger error in the quantity to be calculated.

\section{Cluster Monte Carlo Algorithms}
The Metropolis algorithm makes changes locally one site at a time.
This is the cause of the critical slowing down.  There are algorithms
that change the states by a group of clusters.  The advantage of the
cluster algorithms \cite{Reviews} is that they are faster, not
necessarily in computer time per Monte Carlo step, but in terms of
dynamics.  That is, it has a much smaller value of $z$.

\subsection{Swendsen-Wang algorithm}
The Swendsen-Wang multi-cluster algorithm \cite{SwendsenWang} is
closely related to percolation \cite{Stauffer} based on the work of
Fortuin and Kasteleyn \cite{FortuinKasteleyn}.  One starts with a spin
configuration $\{\sigma\}$ and generates a percolation configuration
based on the spin configuration by the rule described below.  Then the
old spin configuration is forgotten and a new spin configuration
$\{\sigma'\}$ is generated based on the percolation configuration. The
rule is such that the detailed balance is satisfied.  Or more
generally, the transition leaves the equilibrium probability
invariant.  The algorithm goes as follows:
\begin{enumerate}
\item  Start with some arbitrary state $\{\sigma\}$.
\item  Go through each nearest neighbor connection of the lattice,
create a bond between the two neighboring sites $i$ and $j$
with probability $p = 1 - e^{-2J/k_B T}$, but only if the spins
are the same, $\sigma_i = \sigma_j$.  Never put a bond between the
sites if the spin values are different.   We are creating a bond percolation
configuration with probability $p$ on a subset of lattice sites where
the spins are either all pointing up or down.
\item Identify clusters as a set of sites connected by bonds, or isolated
sites.  Two sites are said to be in the same cluster if there is a connected
path of bonds joining them.  Every site has to belong to one of the
clusters.  After the clusters are found, each cluster is assigned a new
Ising spin chosen with equal probability between $+1$ and $-1$.
The old spin values now can be `forgotten'.  And all the
sites in a cluster now take the value of spin assigned to the cluster.
\item One Monte Carlo step is finished.  Repeat~2 for the next
step.
\end{enumerate}

The performance of the algorithm in terms of correlation time in
comparison with Metropolis algorithm is remarkable.  Recall that the
dynamic critical exponent $z$ is about 2 for Metropolis algorithm
almost independent of the dimensionality.  The Swendsen-Wang algorithm
in one dimension gives $z = 0$ ($\tau$ approaches a constant as $T \to
0$). For the two-dimensional Ising model, the dynamic critical
exponent $z$ is less than 0.3 or possibly zero (but $\tau \propto \ln
L$).  In three dimensions it is about 0.5. At and above four
dimensions it is 1.

\subsection{Wolff single-cluster algorithm}
In the Swendsen-Wang algorithm, we generated many clusters and then
flipped these clusters.  Wolff algorithm \cite{Wolff} is a variation on
the way clusters are flipped.  One picks a site at random, and then
generates one single cluster by growing a cluster from the seed.  The
neighbors of the seed site will also belong to the cluster if the
spins are in parallel and a random number is less than $p = 1 -
e^{-2J/k_B T}\!$.  That is, the neighboring site will be in the same
cluster as the seed site with probability $p$ if the spins have the
same sign.  If the spins are different, neighboring site will never
belong to the cluster.  Neighbors of each new site in the cluster are
tested for membership.  This testing of membership is performed on
pair of sites (forming a nearest neighbor bond) not more than once.
The recursive process will eventually terminate.  The spins in the
cluster are turned over with probability 1.  Single cluster algorithm
appears more efficient than the multi-cluster one.

The following is a fairly elegant way of implementing the Wolff
algorithm in C.  The function is the core part which performs a Wolff
single cluster flip.  This function is recursive.  The array for spins
{\tt s[$\;$]}, percolation probability {\tt P}, and coordination
number {\tt Z} (the number of neighbors) are passed globally.  The
first argument {\tt i} of the {\tt flip} function is the site to be
flipped, the second argument {\tt s0} is the spin of the cluster
before flipping. The function {\tt neighbor} returns an array of
neighbors of the current site.  The function {\tt drand48()} returns
a random number unformly distributed between 0 and 1.

\smallbreak
\l{\v`void flip(int i, int s0)`\m{}}
\l{\v`{`\m{}}
\l{\v`   int j, nn[Z];`\m{}}
\smallskip
\l{\v`   s[i] = - s0;`\m{flip the spin immediately}}
\l{\v`   neighbor(i, nn);`\m{find nearest neighbor of {\ninett i}}}
\l{\v`   for(j = 0; j < Z; ++j)`\m{flip the neighbor if}}
\l{\v`      if(s0 == s[nn[j]] && drand48() < P)`\m{spins are equal and}}
\l{\v`         flip(nn[j], s0);`\m{random number is smaller than {\ninett p}}}
\l{\v`}`\m{}}
\smallbreak

\section{Further Development of Cluster Algorithms}
The attractive feature of Swendsen-Wang algorithm stimulated research for
other types of cluster algorithms and its theory.  The
original method works for Ising and Potts models.  It is not clear at
the first sight how one can generalized it for arbitrary model.  The
idea of Ising spin embedding into a model appears to work
well.  This idea is applied to a number of models, including models
with continuous degree of freedom\cite{Wolff,Brower,WangSwendsenKotecky}.
Kandel and coworkers proposed a general scheme for cluster algorithms
\cite{Kandel2,Kandel}, which leads to efficient algorithms for models
with competing interactions \cite{Kandel3,Liang}.  Cluster algorithm
of a different kind (loop algorithm) was proposed to work with vertex
models \cite{Evertz}.

Sokal's idea of introducing auxiliary variables
\cite{Sokal} is another way of generalizing the cluster algorithms and
is drawing attention to statisticians \cite{BesagGreen}.  However, this
method seems not work well in comparison with embedding algorithms
when the system has Ising symmetry and embedding can be implemented.

A large class of problems in statistical mechanics can be now simulated
with cluster algorithms.  Cluster algorithms are used routinely
where traditionally local algorithms are used.  Due to space limitations,
this types of applications will not be reviewed here.

\section{Application in Image Processing}
The Swendsen-Wang algorithm and some generalizations were used in
connection with image processing \cite{BesagGreen}.  Random images
\cite{GrayKayTitterington}, in particular, a class of random fields
called Markov random fields, correspond closely to systems studied in
statistical physics.  Thus the Monte Carlo techniques developed in
statistical mechanics can be readily used here.  There is also the need
to estimate model parameters \cite{PotamianosGoutsias}, this can be
done by Monte Carlo simulation.  The most important application
appears to be in the problems of image restoration\cite{Gray} and
image segmentation\cite{GaudronFrench,Gaudron}.  The statistical
method (or Bayesian inference) in image analysis is relatively new.
The basic problem in Bayesian image restoration is the following.  Let
$X$ be the true image, which is also statistical in nature and
distributed according to $P(X)$, called the prior.  The true scene,
$X$, is not directly observable, and the observed data $Y$ contains
noise.  Given $Y$, one wants to compute the distribution of $X$,
namely, the conditional probability $P(X|Y)$, called posterior.  This
is obtained by noting
$$ P(X|Y) \propto P(Y|X) P(X), \qquad \mbox{for fixed $Y$}. \EQNO$$
Image restoration amounts to find the most probable $X$ by simulating
the probability distribution $P(X|Y)$.  Ising model is assumed for
$P(X)$ in some application; while $P(Y|X)$ is given as independent
Gaussian distribution for $Y$.  From the point of view of statistical
mechanics, $P(X|Y)$ describes a model with inhomogeneous local
magnetic field which is determined by $Y$.

Segmentation is a process of classifying each pixel in an image.  We
can think of it as a special case of restoration problem.  Thus the
two problems share the same mathematical structure.  The problem of
speed of convergence in image segmentation by Swendsen-Wang algorithm
is studied in \cite{GaudronFrench} and \cite{Gaudron}.  The cluster
algorithms work best when the system has a high degree of symmetry.
The two types of problems, image restoration and image segmentation,
all end up involving the simulation of a model with the presence of
complicated magnetic field, which makes cluster algorithms less
effective.  The cluster algorithms and other acceleration algorithms,
e.g., ref. \cite{Novotny}, developed in statistical physics may be
very helpful in this field.  But much more work needs to be done.

\section*{Acknowledgements}
I thank Dr.~S.~Z. Li for inviting me for this talk, and Prof.~K.~V. Mardia
for drawing me the attending of ref \cite{BesagGreen}.

\vglue 2cm
\noindent $^*$For an invited talk at the Second Asian Conference on Computer
Vision, ACCV'95, 5-8 December 1995, Singapore.

\end{document}